\documentclass[aps,prl,twocolumn,superscriptaddress,groupedaddress]{revtex4-2} 
\usepackage[utf8]{inputenc}
\usepackage{graphicx} 
\usepackage{dcolumn} 
\usepackage{bm} 
\usepackage{amssymb}
\usepackage{hyperref} 
\usepackage{amsmath}
\usepackage{enumitem}
\usepackage{nccmath}
\usepackage[usenames,dvipsnames]{xcolor}

\hypersetup{
    pdfencoding=unicode,
	colorlinks=true,
	urlcolor=Maroon,
	linkcolor=RoyalBlue,
	citecolor=OrangeRed,
	pdfdisplaydoctitle=true,
	pdfstartview=FitH,
	linktocpage=true
}
\newcommand\trick[1]{}

\begin{document}
\preprint{YITP-SB-2023-17}

\title{Towards \texorpdfstring{$\alpha'$}{}-finiteness: \texorpdfstring{$q$}{}-deformed open string amplitude}

\author{Yuqi Li}
\affiliation{C.~N.~Yang Institute for Theoretical Physics, State University of New York, Stony Brook, NY 11794-3840}

\author{Hao-Yu Sun}
\affiliation{Weinberg Institute, Department of Physics, The University of Texas at Austin, TX 78712-1192}

\date{\today}

\begin{abstract}
Revisiting the Coon amplitude, a deformation of the Veneziano amplitude with a logarithmic generalization of linear Regge trajectories, we scrutinize its potential origins in a worldsheet theory by proposing a definition of its $q$-deformation through the integral representation of the $q$-beta function. By utilizing $q$-deformed commutation relations and vertex operators, we derive the Coon amplitude within the framework of the dual resonance model. We extend this to the open-string context by $q$-deforming the Lie algebra $\mathfrak{su}(1,1)$, resulting in a well-defined $q$-deformed open superstring amplitude. We further demonstrate that the $q$-prefactor in the Coon amplitude arises naturally from the property of the $q$-integral. Furthermore, we find that two different types of $q$-prefactors, corresponding to different representations of the same scattering amplitude, are essentially the same by leveraging the properties of $q$-numbers. Our findings indicate that the $q$-deformed string amplitude defines a continuous family of amplitudes, illustrating how string amplitudes with a finite $\alpha^\prime$ uniquely flow to the amplitudes of scalar scattering in field theory at energy scale $\Lambda$ as $q$ changes from $1$ to $0$. This happens without the requirement of an $\alpha^\prime$ expansion, presenting a fresh perspective on the connection between string and field theories.
\end{abstract}
\maketitle

\flushbottom

\section{Introduction}
Recently, the Coon amplitude \cite{Coon:1969yw,Baker:1970vxk,Baker:1971jr,Baker:1971zs,Coon:1972qz,Coon:1972te,Coon:1974iu,Baker:1972zv,Arik:1973eb,Arik:1973ve,Baker:1976en,Yu:1972fz,GonzalezMestres:1975ord}, considered as a generalization of the Veneziano amplitude \cite{Veneziano:1968yb} in string scatterings, has regained significant attention. Several remarkable advancements in understanding Coon amplitudes have been achieved, starting from their helping identifying the next-to-leading-order contribution to unitary amplitudes \cite{Caron-Huot:2016icg}, to more direct investigation into their unitarity resulting in the critical dimension \cite{Bhardwaj:2022lbz}, positivity \cite{Chakravarty:2022vrp,Jepsen:2023sia} and uniqueness \cite{Cheung:2022mkw}, their relationship with Virasoro amplitudes \cite{Geiser:2022exp,Geiser:2022icl}, to application of bootstrap techniques \cite{Figueroa:2022onw,Cheung:2023adk}, and even to their interpretation in terms of D-branes \cite{Maldacena:2022ckr}. However, despite early attempts at identifying a string-theoretic origin of these formulas \cite{Coon:1972te,Arik:1973eb,Romans:1988qs,Romans:1989di,Bernard:1989jq,Saito:1990mb,Chaichian:1992hr,Chaichian:1993ft,Jenkovszky:1994qg,Chaichian:1994tr}, a full understanding of whether a worldsheet model of the Coon amplitude exists remains elusive.

As established in pioneering studies \cite{Baker:1970vxk,Baker:1971zs,Coon:1972qz,Coon:1974iu,Arik:1973ve,Baker:1976en,Romans:1988qs,Romans:1989di,Bernard:1989jq,Saito:1990mb}, the four-point Coon amplitude
\begin{equation}\label{Coon_4}
    \mathcal{V}^4_C(\alpha_q(s),\alpha_q(t))=q^{\alpha_q(s)\alpha_q(t)}B_q(-\alpha_q(s),-\alpha_q(t)),
\end{equation}
where $\alpha_q(x):=\ln(1-(1-q)\alpha(x))/\ln q$, and $\alpha(x)$ represents the linear Regge trajectory with respect to the Mandelstam variable $x$, is written as a continuous family of amplitudes parameterized by a single real parameter $q\in[0,1]$. This is achieved with the help of the so-called $q$-beta function \cite{askey1978q,ernst2000history,kac2002quantum,desole2003integral,Ernst2003,annaby2012q,elmonser2015some},
$B_q(\alpha,\beta):=\frac{\Gamma_q(\alpha)\Gamma_q(\beta)}{\Gamma_q(\alpha+\beta)}$ for $\alpha,\beta>0$,
where $q$-gamma function $\Gamma_q(t):=\frac{(1-q)_q^{t-1}}{(1-q)^{t-1}}$, with $(1-q)_q^{t-1}$ being the $q$-analogue of $(1-q)^{t-1}$\footnote{The $q$-analogue of $(x-a)^n$ is the polynomial
\setlength{\belowdisplayskip}{0pt} \setlength{\belowdisplayshortskip}{0pt}
\setlength{\abovedisplayskip}{0pt} \setlength{\abovedisplayshortskip}{0pt}
\begin{equation*}
    \quad(x-a)_q^n:=
    \begin{cases}
        1, \quad\quad\quad\quad\quad\quad\quad\quad&n=0,\\
        (x-a)(x-qa)\dots(x-q^{n-1}a), & n\geq1,
    \end{cases}
\end{equation*}
and can be generalized to any real power $\alpha$ via a new definition $(1+x)_q^\alpha:=\frac{(1+x)_q^\infty}{(1+q^\alpha x)_q^\infty}$ \cite{kac2002quantum}.}. It has a nice integral representation
\begin{eqnarray}\label{q-beta_int}
    B_q(\alpha, \beta)=\int_0^1\mathrm{~d}_q x\,\, x^{\alpha-1}(1-qx)_q^{\beta-1},
\end{eqnarray}
where $\mathrm{d}_q x$ is the $q$-analogue of the ordinary integration measure $\mathrm{d} x$, and $(1-qx)_q^{\beta-1}$ is the $q$-analogue of $(1-x)^{\beta-1}$ \footnote{One can also express $(1-qx)_q^{a-1}$ [the $q$-analogue of $(1-x)^{a-1}$] as the $q$-Pochhammer symbol $(qx;q)_{a-1}$, using the definition $(a;q)_n:=\prod_{k=0}^{n-1}(1-aq^k)$.}. When attempting to view the integral representation of the $q$-beta function as a certain worldsheet integral however, one realizes that $x^{\alpha-1}$ and the $q$-analogue $(1-qx)_q^{\beta-1}$ there cannot be treated as equals, in the same manner as in conventional string theory. Thus, the precise definition of the $q$-deformed string \footnote{For instance, one potential definition of the $q$-deformed string, known as the ``$q$-string'', is outlined in \cite{Romans:1988qs,Romans:1989di,Bernard:1989jq,Saito:1990mb}.} remains unclear. 

To establish a proper definition for the $q$-deformed string, we first need to understand the origins of the two foregoing factors in the integrand in (\ref{q-beta_int}). Noting that $x^{\alpha-1}$ is not $q$-deformed, we will initially assume that it originates from the propagator, and that $(1-qx)_q^{\beta-1}$ is derived from the vertex operators, both in the dual resonance model (DRM). After appropriately $q$-deforming the commutation relations and vertex operators, the Coon amplitude emerges through calculations using DRM. This approach differs from the set-up of \cite{Chaichian:1992hr, Jenkovszky:1994qg}, presenting a novel and distinct perspective on the $q$-deformation of string amplitudes. As a consequence, the prefactor $q^{\alpha_q(s)\alpha_q(t)}$ in the Coon amplitude naturally arises directly from the properties of the polynomials in the integrand, rather than being inserted manually.

In advancing to the open (super)string case \cite{Romans:1988qs,Romans:1989di,Bernard:1989jq}, we choose to $q$-deform the Virasoro subalgebra $\mathfrak{su}(1,1)$ rather than the equivalent $\mathfrak{sl}(2,\mathbb{R})$, because their $q$-deformed versions, defined with $0<q<1$ and $q$ being a root of unity respectively, are not isomorphic \cite{takhtajan1989quantum,Ruegg1993}. By modifying the vertex operators, we are able to derive a well-defined $q$-deformed open superstring amplitude. Interestingly, we discover that both prefactors $q^{\alpha_q(s) \alpha_q(t)}$ and $q^{\alpha_q(s) \alpha_q(t)-\alpha_q(s)-\alpha_q(t)}$ are correct but correspond to different representations of the scattering amplitude. In the DRM calculations, we $q$-deform the commutation relation between the creation ($a_n^{\mu\dagger}$) and annihilation ($a_n^\mu$) operators to $[a_m^\mu, a_n^{\nu \dagger}]_q=g^{\mu \nu} \delta_{m, n}$, using the definition of the $q$-deformed commutator $[A,B]_q:=AB-qBA$ for two operators $A$ and $B$. Whereas in the superstring calculations, we employ $\left[\alpha^\mu_m, \alpha^\nu_n\right]_q=\left[m\right]_q g^{\mu \nu} \delta_{m+n, 0}$ with $[m]_q$ being a $q$-number, the $q$-analogue of an integer, defined as $[m]_q:=(q^m-1)/(q-1)$. Without changing the definitions of the $q$-analogues of logarithm and exponential functions, we find that both commutation relations yield consistent results.
By interpreting $q$ as a function of $\alpha^\prime$, we propose that the $q$-deformed string amplitude can be viewed as a continuous family that uniquely determines the transition from string theory, characterized by a fixed finite $\alpha^\prime$, to scattering of scalars in field theory, determined by a certain energy scale $\Lambda$, without any necessity of the expansion in terms of $\alpha^\prime$. Finally, we discuss the possible future directions based on our results.

\section{\texorpdfstring{$q$}{}-deformed operator formalism and Coon amplitude}
One of several systematic investigations into Coon amplitudes in the early days was achieved by using the DRM in \cite{Romans:1988qs}. Not based on Lagrangian density, DRM perturbatively describes hadrons, with a set of axioms including Lorentz invariance, unitarity, $T,C,P$ invariance, crossing symmetry, analyticity with only poles in the complex plane of the Mandelstam variables and in that of angular momenta (``Regge behavior''), among others. For a complete list for empirical applications and approximations, see \cite{Schwarz:1973yz,Scherk:1974jj}. It differs from conventional field theories due to an infinite number of resonances even at the lowest order in the real coupling constant.

Now consider the simplest vertex operators corresponding to Veneziano amplitudes and perform the $q$-deformation on the operators. We obtain the $q$-deformed vertex operator with momentum $k$ as follows \cite{Schwarz:1973yz},
\begin{eqnarray}\label{vertex}
V(k)&=&\exp_q \left(\sqrt{2 \alpha^{\prime}} k \cdot \sum_{n=1}^{\infty} \frac{a_n^{\dagger}}{\sqrt{[n]_q}}\right)\nonumber\\
&&\times\exp_q \left(-\sqrt{2 \alpha^{\prime}} k \cdot \sum_{n=1}^{\infty} \frac{a_n}{\sqrt{[n]_q}}\right),
\end{eqnarray}
with the $q$-exponential defined as
$\exp_q(x):=\sum_{k=0}^{\infty} z^k/[k]_q!$. Here, we adopt our commutation relations as
\vspace{-0.1cm}
\begin{equation}
[a_m^\mu, a_n^{\nu \dagger}]_q=g^{\mu \nu} \delta_{m, n},\quad m, n=1,2,3,\dots,
\end{equation}
\vspace{-0.2cm}

\hspace{-0.35cm}where $g^{\mu\nu}$ is the inverse metric of the flat spacetime in Lorentzian signature. There is no a priori dimension of the spacetime \footnote{Although a critical dimension $d=26$ of the DRM can be obtained by considering the Pomeron-like pole in dual loop diagrams \cite{Goddard:1972iy}, or by considering the spectrum-generating algebra \cite{Brower:1972wj}.}.
After verifying the completeness of the Bargmann--Fock space, as explored in prior work \cite{Bracken:1990dw}, we then proceed to define the corresponding propagator as 
\begin{eqnarray}
D(s)=\int_0^1 x^{R-\alpha(s)-1}(1-qx)_q^{\alpha_0-1} \mathrm{~d}_q x,
\end{eqnarray}
with the $q$-analogue of the number operator defined as
\begin{eqnarray}
R=\sum_{n=1}^{\infty} [n]_q\, a_n^{\dagger} \cdot a_n.
\end{eqnarray}
This leads to the $q$-deformed 4-point Veneziano amplitude, which is expressed as
\begin{eqnarray}
\mathcal{V}_q^4=\left\langle 0_q\left|V\left(p_2\right) D\left(s\right) V\left(p_3\right) \right| 0_q\right\rangle,
\end{eqnarray}
with $a_n^{\mu}|0_q\rangle=0$ for $n>0$ as in \cite{Chaichian:1992hr}. On substituting the expressions of the corresponding vertex operators in (\ref{vertex}), one obtains:
\begin{widetext}
 \begin{eqnarray}
\mathcal{V}_q^4(s,t)&=&\int_0^1 \mathrm{~d}_qx\,\, x^{-\alpha(s)-1}(1-\textcolor{orange}{q}x)_q^{\alpha_0-1}\nonumber\\
&&\times\left\langle 0_q\left|\exp_q \left(-\sqrt{2 \alpha^{\prime}} p_2 \cdot \sum_{n=1}^{\infty} \frac{a_n}{\sqrt{[n]_q}}\right) x^R \exp_q \left(\sqrt{2 \alpha^{\prime}} p_3 \cdot \sum_{n=1}^{\infty} \frac{a_n^{\dagger}}{\sqrt{[n]_q}}\right)\right| 0_q\right\rangle\nonumber\\
&=&\int_0^1 \mathrm{~d}_qx\,\, x^{-\alpha(s)-1}(1-\textcolor{orange}{q}x)_q^{\alpha_0-1}\exp_q \left(-2 \alpha^{\prime} p_2 \cdot p_3 \sum_{n=1}^{\infty} \frac{x^n}{[n]_q}\right)\nonumber\\
&=&\int_0^1 \mathrm{~d}_qx\,\, x^{-\alpha(s)-1}(1-\textcolor{ForestGreen}{q^{\alpha(t)}}\textcolor{orange}{q} x)_q^{-\alpha(t)-1}\label{q-deformed}\\
&=&\textcolor{ForestGreen}{q^{\alpha(s)\alpha(t)}}\int_0^1 \mathrm{~d}_qx\,\, x^{-\alpha(s)-1}(1-\textcolor{orange}{q} x)_q^{-\alpha(t)-1}.
\end{eqnarray}
\end{widetext}
Here, we use the relation $e^Ae^B=e^Be^Ae^{[A,B]_q}$ when $[A,B]_q$ is a $c$-number and in the last line of the calculation, we perform the substitution $\textcolor{ForestGreen}{q^{\alpha(t)}}x\rightarrow x$. In step (\ref{q-deformed}), we utilized the following useful identity \cite{kac2002quantum,Ernst2003}:
\begin{eqnarray}\label{q-identity}
\frac{(a+b)_q^m}{(a+b)_q^n}=(a+q^nb)_q^{m-n},\quad m,n\geq0.
\end{eqnarray}
In our calculation, we employed a more generalized definition $(a+b)_q^m:=\prod_{k=0}^{m-1}(a+q^kb)$, which immediately gives us $(1-qx)_q^{m-1}=(1-x)_q^{m}/(1-x)_q$ as already indicated earlier. This identity will be used repeatedly in our subsequent calculations. 
The calculations so far have been performed in the kinematic space, yielding what we refer to as the $q$-Veneziano amplitude. However, this is not yet the Coon amplitude. The Coon amplitude is defined in the $q$-space, with its coordinates given by the logarithmic Regge trajectory, which is mapped from its linear Regge trajectory in the kinematic space \cite{Romans:1988qs,Romans:1989di}. For any Regge trajectory $\alpha(x)$, we define $\alpha_q(x):=\frac{\ln[1-(1-q)\alpha(x)]}{\ln q}$. Finally, we obtain the Coon amplitude as shown in (\ref{Coon_4}):
\begin{eqnarray}\label{Coon_amplitude}
\mathcal{V}_C^4(s,t)&=&q^{\alpha_q(s)\alpha_q(t)}\int_0^1 \mathrm{~d}_qx\,\, x^{-\alpha_q(s)-1}(1-qx)_q^{-\alpha_q(t)-1}\nonumber\\
&=&q^{\alpha_q(s)\alpha_q(t)}B_q(-\alpha_q(s),-\alpha_q(t)).
\end{eqnarray}
\section{From \texorpdfstring{$q$}{}-deformed \texorpdfstring{$\mathfrak{su}(1,1)$}{} to \texorpdfstring{$q$}{}-strings}
Upon appropriately normalizing $\alpha_0$ in the linear Regge trajectory, the calculation using the DRM can be interpreted as the scattering amplitude of tachyons. The integral representation of four-tachyon scattering raises an intriguing question about the existence of a corresponding expression for the scattering of excited string states in terms of $q$-analogues of gamma functions. Building on the groundwork laid by previous research \cite{Romans:1988qs,Romans:1989di,Bernard:1989jq}, our endeavor aims to construct a well-defined $q$-deformed string ($q$-string) theory. 

To start, let us first focus on the tree-level scattering amplitude of four massless open superstrings with $\mathfrak{su}(1,1)$ as the subalgebra of the Virasoro algebra. In the conventional calculations of conformal field theory (CFT), we set $(0,1,\infty)$ as the three punctures on the Riemann surface fixed by the global conformal group $\textrm{SU}(1,1)$ and the fourth puncture is located at $x$ on the real axis. In the $q$-deformed case, since $[0]_q=0$ and $[1]_q=1$ remain unchanged under the $q$-deformation, we can still use the same three punctures fixed by the group corresponding to the $q$-deformed $\mathfrak{su}(1,1)$ algebra. This is possible as long as the terms corresponding to the $q$-analogue of $\infty$ in the calculation cancel out to yield some polynomials of degree $0$. This is exactly the same case as in conventional string amplitude calculations \cite{Zwiebach:2004tj}. In contrast to the traditional commutation relations used in the DRM calculations above, we will use the $q$-analogue of the corresponding notations often used in string amplitude calculations to demonstrate that both commutations used in DRM and string theory are consistent sets of commutation relations:
\begin{equation}
    \left[\alpha^\mu_m, \alpha^\nu_n\right]_q=\left[m\right]_q g^{\mu \nu} \delta_{m+n, 0}.\label{eq:comm_string}
\end{equation}
The key modifications in the calculations of the $q$-deformed string amplitude are listed as follows:
\begin{itemize}
\setlength\itemsep{0.3em}
    \item \textbf{The mode expansion of the string field $X^\mu$ is transformed into its $q$-analogue.} For instance, if we assume that both the left and right movers (corresponding to $\alpha^\mu_n$ and $\tilde{\alpha}^\mu_n$, respectively) are deformed under the same $q$ \footnote{In general, they are not required to be deformed in the same way, even if the level-matching condition is satisfied. We will discuss this further in our future work on the $q$-deformation of closed strings.}, we obtain the following mode expansion:
    \begin{eqnarray}\label{mode_expansion}
    X^\mu(z, \bar{z})&=&\mathit{x}^\mu-i \mathit{p}^\mu\left[ 2\alpha^\prime \ln_q (z \bar{z})\right]\nonumber\\
    &&\hspace{-0.2cm}+i\sqrt{2\alpha^\prime} \sum_{n \neq 0} \frac{1}{[n]_q}\left(\alpha^\mu_n z^{-n}+\tilde{\alpha}^\mu_n \bar{z}^{-n}\right),
    \end{eqnarray}
    with $\ln_q(\cdot)$ being the inverse function of $\exp_q(\cdot)$.
    \item \textbf{The $q$-deformed commutation relations implies $[p^\mu,x^\nu]_q=-ig^{\mu\nu}$ and the $q$-Wick's theorem.} Note that the $q$-deformation only affects the indices of the worldsheet coordinates and leaves the spacetime indices untouched. The $q$-deformation necessitates the $q$-Wick's theorem, and it enforces the normal ordering of two bosonic operators such that $:bb^\dagger:=qb^\dagger b$, as described in \cite{2007JPhA...40.8393M}.
    \item \textbf{The partial derivatives acting on worldsheet coordinates are $q$-deformed.} This essentially involves the substitution of the corresponding $q$-analogue of differential operators: $\frac{\partial}{\partial z}\rightarrow\left(\frac{\partial}{\partial z}\right)_q=\frac{1}{z}\left[z\frac{\partial}{\partial z}\right]_q$ and $\frac{\partial}{\partial \bar{z}}\rightarrow\left(\frac{\partial}{\partial \bar{z}}\right)_q=\frac{1}{\bar{z}}\left[\bar{z}\frac{\partial}{\partial \bar{z}}\right]_q$.
    
    \item \textbf{The string vertex operators maintain the same form with the aforementioned modified $q$-analogues, and the operator product expansions (OPEs) remain unchanged.} This point is straightforward since the OPEs used in the amplitude calculations only involve terms of order $1/(z-w)$, which are not affected by the $q$-analogue.
    \item With all the substitutions mentioned above, the kinematic factor of the $q$-deformed string amplitudes remains unchanged.
\end{itemize}
The modifications mentioned above enable the computation of the $q$-deformed string amplitudes using the standard techniques found in conventional string amplitude calculations, as demonstrated in \cite{Polchinski:1998rq}. After a series of straightforward calculations,  we find that by setting $(z_1,z_2,z_3,z_4)=(0,x,1,\infty)$, the color-stripped amplitude $\mathcal{A}_q(1,2,3,4)$ of four massless open superstrings at tree level can be expressed as follows:
\begin{equation}
\begin{aligned}
&\mathcal{A}^4_q(s,t)=-2 \alpha^{\prime} \int_0^1 \mathrm{d}_q x\left(\frac{N_s}{x}-\frac{N_t}{1-x}\right)\\
&\times\exp_q{\left[-2 \alpha^{\prime} s_{12}\ln_q(x-0)\right]}\exp_q{\left[-2 \alpha^{\prime} s_{23}\ln_q(1-x)\right]}\\
=&-2 \alpha^{\prime} \int_0^1 \mathrm{d}_q x\left(\frac{N_s}{x}-\frac{N_t}{1-x}\right)x^{-\alpha^{\prime} s}\left(1-q^{\alpha^{\prime} t}x\right)_q^{-\alpha^{\prime} t}.
\end{aligned}
\end{equation}
Here, $s=2s_{12}=2k_1\cdot k_2$ and $t=2s_{23}=2k_2\cdot k_3$ represent the Mandelstam variables, while $N_s$ and $N_t$ correspond to the Bern--Carrasco--Johansson (BCJ) numerators \cite{Mafra:2022wml}. Note that we utilized the definition of the $q$-analogue such that $(x-0)_q^a=x^a$, which reverts to a non-deformed expression. Now, let us map from the kinematic space to the $q$-space as introduced in the previous section by setting $\alpha(s)=\alpha^\prime s$ and $\alpha(t)=\alpha^\prime t$. In the $q$-coordinate, for the part involving $N_s$, we have:
\begin{widetext}
\begin{equation}
\begin{aligned}
&N_s(2\alpha^\prime)\int_{0}^{1} \mathrm{d}_q x\,\, x^{-\alpha_q(s)-1}(1-\textcolor{ForestGreen}{q^{\alpha(t)-1}} \textcolor{orange}{q}x)_q^{-\alpha_q(t)}\\
=&\textcolor{ForestGreen}{q^{\alpha_q(s) \alpha_q(t)-\alpha_q(s)}}\frac{N_s}{s}([\alpha_q(s)]_q)\int_{0}^{1} \mathrm{d}_q x\,\, x^{-\alpha_q(s)-1}(1-\textcolor{orange}{q}x)_q^{-\alpha_q(t)}\\
=&\textcolor{ForestGreen}{q^{\alpha_q(s) \alpha_q(t)-\alpha_q(s)}}\frac{N_s}{s}(-q^{\alpha_q(s)}[-\alpha_q(s)]_q)\int_{0}^{1} \mathrm{d}_q x\,\, x^{-\alpha_q(s)-1}(1-\textcolor{orange}{q}x)_q^{-\alpha_q(t)}\\
=&-\textcolor{ForestGreen}{q^{\alpha_q(s) \alpha_q(t)}}\frac{\Gamma_q(1-\alpha_q(s))\Gamma_q(1-\alpha_q(t))}{\Gamma_q(1-\alpha_q(s)-\alpha_q(t))}\left[\frac{N_s}{s}\right].\\
\end{aligned}
\end{equation}
\end{widetext}
Here, we used $\Gamma\left(1-\alpha_q(s)\right)=\left[-\alpha_q(s)\right]_q\Gamma_q(-\alpha_q(s))$ and $[-a]_q=-q^{-a}[a]_q$. 
The calculation of the part corresponding to $N_t$ is similar. After summing up the two parts and setting $A_4=A_{SYM}(1,2,3,4):=\frac{N_s}{s}-\frac{N_t}{t}$ and $q(s,t)=q^{\alpha_q(s)\alpha_q(t)}$, we obtain the $q$-deformed four-point amplitude:
\begin{eqnarray}
\mathcal{A}^4_q(s,t)=\textcolor{ForestGreen}{q(s,t)}A_4\frac{\Gamma_q(1-\alpha_q(s))\Gamma_q(1-\alpha_q(t))}{\Gamma_q(1-\alpha_q(s)-\alpha_q(t))}.
\end{eqnarray}
If we apply the property of $q$-gamma function backwards and set $\tilde{q}(s,t)=q^{\alpha_q(s) \alpha_q(t)-\alpha_q(s)-\alpha_q(t)}$, we can get the other form of the amplitude with the kinematic factors $K_4$:
\begin{equation}
\begin{aligned}
\mathcal{A}^4_q&=\textcolor{ForestGreen}{\tilde{q}(s,t)}\left[\frac{K_4}{st}\right]\frac{\Gamma_q(-\alpha_q(s))\Gamma_q(-\alpha_q(t))}{\Gamma_q(1-\alpha_q(s)-\alpha_q(t))},\\
\end{aligned}
\end{equation}
which agrees with the result in \cite{Geiser:2022icl}. Furthermore, our method can be straightforwardly extended to higher-point amplitudes by carefully handling the $q$-prefactors with respect to the kinematic factors. For 5- and 6-point amplitudes, we observe similar results to those in \cite{Romans:1988qs}, and the appearance of the $q$-prefactors emerges naturally. The $q$-deformed 5-point amplitude of tachyons in the kinematic space can be expressed as follows:
\begin{widetext}
\begin{equation}
\begin{aligned}
\mathcal{V}_q^5(s_{12},s_{23},s_{34},s_{45},s_{51})&=\textcolor{ForestGreen}{q^{\alpha(s_{34})\alpha(s_{45})+\alpha(s_{12})\alpha(s_{23})}}\int_0^1 \mathrm{d}_q y\,\, y^{-\alpha(s_{45})-1}(1-\textcolor{orange}{q}y)_q^{-\alpha(s_{34})-1}\\
&\times\int_0^1 \mathrm{d}_q x\,\, x^{-\alpha(s_{12})-1}(1-\textcolor{orange}{q}x)_q^{-\alpha(s_{23})-1}(1-\textcolor{orange}{q^{-\alpha(s_{23})-\alpha(s_{34})}}xy)_q^{-\alpha(s_{51})+\alpha(s_{23})+\alpha(s_{34})}\\
&=\textcolor{ForestGreen}{q^{\alpha(s_{34})\alpha(s_{45})+\alpha(s_{12})\alpha(s_{23})}}B_q\left(-\alpha\left(s_{34}\right),-\alpha\left(s_{45}\right)\right) B_q\left(-\alpha\left(s_{12}\right),-\alpha\left(s_{23}\right)\right)\\
&\times\,_3\phi_2\left(\begin{array}{c}
    q^{-\alpha\left(s_{12}\right)}, q^{-\alpha\left(s_{45}\right)}, q^{\alpha\left(s_{51}\right)-\alpha\left(s_{23}\right)-\alpha\left(s_{34}\right)} \\
    q^{-\alpha\left(s_{12}\right)-\alpha\left(s_{23}\right)}, q^{-\alpha\left(s_{34}\right)-\alpha\left(s_{45}\right)}
    \end{array}; q, q^{-\alpha\left(s_{51}\right)}\right).
\end{aligned}
\end{equation}
\end{widetext}
Here, we consider 5 points located at $(z_1,z_2,z_3,z_4,z_5)=(0, x, y, 1, \infty)$ and define $\alpha(s_{ij})$ as the linear Regge trajectory in terms of the Mandelstam variables $s_{ij}$. The expression $_3\phi_2$ denotes the unilateral basic hypergeometric series \footnote{The unilateral basic hypergeometric series is defined as 
\begin{equation*}
\begin{aligned}
\quad\quad_j&\phi_k\left(\begin{array}{c}
    a_1, \dots, a_j \\
    b_1,\dots,b_k
    \end{array}; q, z\right)\\&:=\sum_{n=0}^\infty\frac{(a_1;q)_n\cdots(a_j;q)_n}{(b_1;q)_n\cdots(b_k;q)_n(q;q)_n}\left((-1)^nq^{{n\choose2}}\right)^{1+k-j}z^n,
    \end{aligned}\end{equation*} 
    see e.g., \cite{kac2002quantum}.
    It is the $q$-analogue of the hypergeometric series because \begin{equation*}
    \begin{aligned}\lim_{q\rightarrow1} \,\hspace{-0.1cm}_j\phi_k&\left(\begin{array}{c}
    q^{a_1}, \dots, q^{a_j} \\
    q^{b_1},\dots,q^{b_k}
    \end{array}; q, (q-1)^{1+k-j}z\right)\\&\quad\quad\quad\quad\quad\quad\quad\quad=\,\hspace{-0.1cm} _j F_k\left(\begin{array}{c}
    a_1, \dots, a_j \\
    b_1,\dots,b_k
    \end{array}; z\right).\end{aligned}\end{equation*}\protect\trick.} (also known as $q$-hypergeometric series) \cite{gasper2004basic,andrews1974applications,askey1980some}.
Indeed, the last term in the integrand, derived from the properties of Mandelstam variables, is valid only for the linear Regge trajectories, and it is specific to the kinematic space. This term is crucial for obtaining the correct $q$-deformed 5-point amplitude within the given framework. However, when transitioning to the $q$-space, where the Coon amplitude is defined, this term may require further modifications or adaptations to ensure the consistency of the Coon amplitude in the $q$-space. We will leave the details of the calculations of $N$-point amplitude to our future work.

\section{The relationship between \texorpdfstring{$q$}{Lg} and \texorpdfstring{$\alpha^\prime$}{Lg} at low energy}
As already shown in \cite{Figueroa:2022onw,Chakravarty:2022vrp,Bhardwaj:2022lbz,Cheung:2022mkw,Cheung:2023adk} and references therein, the Coon amplitude in $q$-space interpolates between the field-theoretic results and the stringy results at the endpoints $q=0$ and $q=1$, respectively. In the field-theoretic limit as $q\rightarrow0$, the non-contact part of the Coon amplitude has simple poles with an overall factor of the stringy parameter $\alpha^\prime$. This can be represented as follows:
\begin{equation}
\lim _{q \rightarrow 0} V^4_{q}(s,t)=-\frac{1}{\alpha^\prime(s-m^2)}-\frac{1}{\alpha^\prime(t-m^2)}+1.
\end{equation}
In kinematic space, the $q$-deformed Veneziano amplitude allows us to take a step further in our exploration of whether, for fixed and finite arbitrary $\alpha^\prime$, the dependence on $\alpha^\prime$ can be absorbed into a dimensional prefactor $\Lambda$ so that the resulting four-point scattering in the $q\rightarrow0$ limit will be independent of the stringy characteristic $\alpha^\prime$. Note that, similar to the standard gamma function, the $q$-gamma function $\Gamma_q(z)$ also has a pole at $z=0$ \cite{Berkooz:2018jqr}, \footnote{Using the expression $\Gamma_q(z+1)=[z]_q\Gamma_q(z)$, we can perform a direct calculation which demonstrates that the pole is located at $z=0$, with the residue being $\frac{q-1}{\ln q}$.}
\begin{eqnarray*}
    \Gamma_q(z)\approx \frac{q-1}{\ln q}\frac{1}{z}+\mathcal{O}(1).
\end{eqnarray*}
Thus, we can expand the $q$-deformed Veneziano amplitude around the poles and keep the non-contact leading order terms:
\begin{equation}
\begin{aligned}
V_q^4(s,t)_{\texttt{non}}&\sim\left[\frac{1-q}{\ln q}\frac{1}{\alpha^\prime}\right]\left(\frac{1}{s-m^2}+\frac{1}{t-m^2}\right).\\
\end{aligned}
\end{equation}
Next, let us absorb the overall factor into $\Lambda$ and set $\Lambda=\frac{q-1}{\ln q}\frac{1}{\alpha^\prime}$. Upon taking the small $q$ limit, the leading contribution yields $q\approx\exp{(-\frac{1}{\alpha'\Lambda})}$. Before proceeding, it is essential to carry out some consistency checks of different limits:
\begin{itemize}
\setlength\itemsep{0.3em}
\item \textbf{$\alpha'\rightarrow0$ limit:} In the standard string theory calculations, $\alpha^\prime\rightarrow0$ limit of the string amplitude calculations corresponds to the field theory result. In our case, $\alpha^\prime\rightarrow0$ implies $\exp{(-\frac{1}{\alpha'\Lambda})}\rightarrow0$, which corresponds to the scalar field theory result.
\item \textbf{$\alpha'\not\approx0$ and fixed:} Let us set $\Lambda=\lambda m_0^2$, with $\lambda\in\mathbb{R^+}$ and $m_0^2=-m_T^2$, where $m_T^2$ denotes the squared mass of the tachyon. In this case, the limit $q\rightarrow0$ implies $\alpha^\prime m_T^2\rightarrow0$. This is consistent as $\alpha^\prime m_T^2=\frac{D-2}{2}\zeta_q(-1)$ is linearly suppressed in the $q\rightarrow0$ limit. Here, $\zeta_q(z)$ represents the $q$-deformed Riemann zeta function.
\end{itemize}
Now, let us fix the energy scale of the scalar field theory to be $\Lambda=\lambda m_0^2$ and define
\begin{eqnarray}
q_0=\exp\left(-\frac{1}{\alpha^\prime m_0^2}\frac{1}{\lambda}\right),\quad \lambda\in\mathbb{R^+}.
\end{eqnarray}
All remaining terms can then be absorbed into the overall factor $\mathcal{R}(\lambda,m_0^2)$, such that $q=q_0\mathcal{R}(\lambda,m_0^2)$. Here, a crucial observation is that when $q=q'$, $\Gamma_q(x)=\Gamma_{q'}(x)$ and $\Gamma_q(x)<\Gamma_{q'}(x)$ if $q<q'$. Therefore, as $q$ flows from $1$ to $0$, it uniquely defines a transition from a string theory characterized by a finite $\alpha^\prime$ to a field theory with a fixed energy scale $\Lambda$. It becomes evident that the $q$-deformation can be interpreted as a continuous family that demonstrates how string theory flows into field theory without the need for a perturbative expansion of $\alpha^\prime$. Instead, this transition is facilitated through a complicated function that maps $\alpha^\prime$ to $q$.
\section{Discussions and outlook}

In summary, our calculation proposes a definition for the Coon amplitude, with the appropriate prefactor arising directly from the $q$-integral. We extend this approach to the open string amplitude through a $q$-deformation of the Lie algebra $\mathfrak{su}(1,1)$, which yields a well-defined $q$-deformed open superstring amplitude. Our results indicate that the $q$-deformation gives rise to a continuous family of amplitudes, offering a novel perspective on the relationship between string theory and field theory.

Before moving to the outlook of future directions, we will discuss more on the details of our findings. Noticing that $\exp_q(a\ln b)\neq\exp_q (a\ln_qb)=b_q^a$, with $b_q^a$ being the $q$-analogue of $b^a$ as usual, we advance our understanding beyond the work of \cite{Chaichian:1992hr}, proposing a more canonical definition of the $q$-deformation of the mode expansion of the string field, as given in (\ref{mode_expansion}). It is only with this refined definition that the correct form of the Coon amplitude emerges. When we associate the deformation parameter $q$ with $\alpha^\prime$, the curve in the $(q,D)$ space proposed by \cite{Bhardwaj:2022lbz} can be reinterpreted as a diagram in the $(\alpha^\prime,D)$ space, which illustrates the point at which the amplitude consistently maintains unitarity as $\alpha^\prime$ approaches zero. It is intriguing that, after suitably defining our $\lambda m_0^2$, we obtain a behavior similar to that observed in \cite{Maldacena:2022ckr}: $q\sim e^{-\frac{1}{\alpha^\prime m_\infty^2}}$. This resemblance is particularly striking considering that our results are derived in the low energy limit.

Let us wrap up our exploration by the outlook of the future directions:
Our work has provided a potential new perspective on the $q$-Virasoro algebra. Based on the previous studies \cite{Daskaloyannis:1991ku,Chaichian:1990rt,Hiro-Oka:1990dpn,Chaichian:1996qy,Saito:1990mb,Sato:1991sj,Hounkonnou:2018pxg}, the $q$-deformation of the Virasoro algebra and the mass spectrum can be further explored. It would be very interesting to see how closed string theory can be formulated based on our findings in the open string scenario, including the potential for a $q$-analogue of double copy relations. If we interpret our deformed algebra as a deformation of the universal enveloping algebra, then the Hopf and braiding structures will naturally arise. Thus, investigating the Hopf structures of the $q$-deformed string amplitudes would be an exciting avenue for further research.

\onecolumngrid

\bigskip

\paragraph{Acknowledgements}
We express our gratitude to Piotr Tourkine for enlightening discussions and incisive comments on this work. We also thank Peng Zhao for meticulously reviewing the initial draft and providing useful suggestions. Special thanks go to Marcus Spradlin for bringing Romans' work to our attention. We thank Martin Ro\v cek and Mykola Dedushenko for valuable discussions as well. 
Y.L. is supported, in part, by NSF grant PHY-22-15093. Y.L. would like to express special thanks to Nima Arkani-Hamed, Lorenz Eberhardt, and Sebastian Mizera for their hospitality at the Institute for Advanced Study, and also extended the gratitude to the workshop ``\href{https://www.ias.edu/sns/string-amplitudes-finite}{String Amplitudes at Finite $\alpha^\prime$}", where the initial ideas for this project were conceived.
H.-Y.S. is supported, in part, by a grant from the Simons Foundation (Grant 651440, AK). H.-Y.S. thanks the hospitality of the Aspen Center for Physics, which is supported by National Science Foundation grant PHY-2210452.
\medskip
\twocolumngrid

\nocite{*}

\bibliography{ref}

\end{document}